\documentclass[12pt,onecolumn]{article}

\usepackage[a4paper,margin=0.75in]{geometry}
\usepackage{times}
\usepackage{ktbox}
\usepackage{ktorcid}

\usepackage{siunitx}      % for \SI and \percent
\usepackage{subcaption}   % for subtable
\usepackage{booktabs}     % for \toprule \midrule \bottomrule

\usepackage{fvextra}     % enhanced verbatim
\usepackage{inconsolata} % or newtxtt, but only for codeblock

\DefineVerbatimEnvironment{codeblock}{Verbatim}{
  fontsize=\footnotesize,
  fontfamily=zi4,          % use Inconsolata only here
  breaklines=true,
  breakanywhere=true,
  breaksymbolleft=\raisebox{0.2ex}{\tiny$\hookleftarrow$},
  numbers=none,
  xleftmargin=0.6em,
  xrightmargin=0pt,
  frame=none,
  samepage=true
}

\usepackage{hyperref}      %% Enables \autoref
\hypersetup{
  colorlinks=true,
  linkcolor=blue,
  citecolor=blue,
  filecolor=magenta,
  urlcolor=cyan,
  breaklinks=true,
}

\usepackage{cite}   % optional, safe, widely available

\usetikzlibrary{arrows.meta}
\usetikzlibrary{positioning}

\newenvironment{keywords}{
  \begin{quote}\small\textbf{Keywords:}%
}{\end{quote}}

\date{} % suppress date

\begin{document}
  %% Title
\title{KTBox: A Modular LaTeX Framework for Semantic Color, Structured Highlighting, and Scholarly Communication}

% KTBox: Semantic Styling and Structured Highlighting for Scientific Authoring
% KTBox Framework: Modular Semantic Color and Structured Boxes for Scholarly Writing

  %% Authors
\author{%
  Bhaskar Mangal\orcidicon{0000-0002-8126-3528}\thanks{B. Mangal (Corresponding Author) is with the Department of CSIS, Birla Institute of Technology and Science (BITS) Pilani, India, and also with C.E. Info Systems Ltd. (MapmyIndia), New Delhi, India. E-mail: \texttt{p20210473@pilani.bits-pilani.ac.in}, \texttt{mangal@mapmyindia.com}.}%
  \and
  Ashutosh Bhatia\orcidicon{0000-0002-3576-0275}\thanks{A. Bhatia, Y. Sharma, and K. Tiwari are with the Department of CSIS, BITS Pilani, India. E-mail: \texttt{\{ashutosh.bhatia, yash, kamlesh.tiwari\}@pilani.bits-pilani.ac.in}.}%
  \and
  Yashvardhan Sharma%
  \and
  Kamlesh Tiwari%
  \and
  Rashmi Verma\thanks{R. Verma is with C.E. Info Systems Ltd. (MapmyIndia), New Delhi, India. E-mail: \texttt{rashmi@mapmyindia.com}.}%
}

  \maketitle

  %% Abstract
  \begin{abstract}
The communication of technical insight in scientific manuscripts often relies on ad-hoc formatting choices, resulting in inconsistent visual emphasis and limited portability across document classes. This paper introduces \texttt{ktbox}, a modular \LaTeX{} framework that unifies semantic color palettes, structured highlight boxes, taxonomy trees, and author metadata utilities into a coherent system for scholarly writing. The framework is distributed as a set of lightweight, namespaced components: \texttt{ktcolor.sty} for semantic palettes, \texttt{ktbox.sty} for structured highlight and takeaway environments, \texttt{ktlrtree.sty} for taxonomy trees with fusion and auxiliary annotations, and \texttt{ktorcid.sty} for ORCID-linked author metadata. Each component is independently usable yet interoperable, ensuring compatibility with major templates such as \texttt{IEEEtran}, \texttt{acmart}, \texttt{iclr conference}, and \texttt{beamer}. Key features include auto-numbered takeaway boxes, wide-format highlights, flexible taxonomy tree visualizations, and multi-column layouts supporting embedded tables, enumerations, and code blocks. By adopting a clear separation of concerns and enforcing a consistent naming convention under the \texttt{kt} namespace, the framework transforms visual styling from cosmetic add-ons into reproducible, extensible building blocks of scientific communication, improving clarity, portability, and authoring efficiency across articles, posters, and presentations.

  \end{abstract}


  %% Introduction
  \section{Introduction}
  \label{sec:introduction}
Scientific communication has expanded beyond the traditional written manuscript into diverse media, including journals, conference proceedings, academic posters, and digital presentations. Across these formats, clarity of exposition and the ability to emphasize core findings remain essential. Although \LaTeX{} has long been the standard for scholarly publishing, existing class and template files focus primarily on layout compliance rather than mechanisms for structured emphasis or semantic visualization. Authors are therefore left with improvised methods to highlight contributions or depict conceptual hierarchies, often through ad-hoc colored boxes, custom diagrams, or informal annotations that lack consistency, portability, and reproducibility.

This situation creates significant limitations for effective knowledge dissemination. Readers are required to parse dense technical text without consistent visual cues for prioritization. Authors expend effort duplicating or modifying formatting code that frequently breaks when reused across venues. Editors and reviewers lack standardized mechanisms for locating central contributions, findings, or limitations. The problem extends beyond manuscripts: academic posters require compact and visually distinct summaries, beamer presentations demand structured elements that capture attention and facilitate audience recall, and surveys increasingly rely on taxonomic trees to organize rapidly expanding research landscapes. The absence of a unified framework for semantic highlighting and taxonomy visualization thus constrains scientific communication at multiple levels.

The \texttt{ktbox} framework addresses this gap by introducing a modular, reusable, and template-agnostic system for semantic styling. Implemented as a suite of namespaced packages, it separates stylistic design from logical structure, ensuring portability across article, poster, survey, and presentation formats. \texttt{ktcolor.sty} defines light and dark semantic palettes, \texttt{ktbox.sty} provides reusable environments for numbered takeaways and thematic highlights, \texttt{ktlrtree.sty} introduces extensible left-to-right taxonomy trees with optional fusion and auxiliary annotations, and \texttt{ktorcid.sty} integrates author metadata through ORCID utilities. Together, these components provide a principled foundation for embedding pedagogical and communicative clarity directly into scholarly documents. The framework is designed not merely as a convenience tool but as a technical contribution to the infrastructure of scientific writing, enhancing accessibility, reproducibility, and long-term impact.

  %% Framework
  \section{Framework}
  \label{sec:framework}
The \texttt{ktbox} framework is composed of three interlinked style files: \texttt{ktbox.sty}, \texttt{ktcolor.sty}, and \texttt{ktorcid.sty}. Each file provides a distinct functionality while maintaining modularity. The design principle is to ensure clear separation of concerns: logical environments for highlight boxes, thematic palettes for stylistic consistency, and author metadata integration. This modularity allows compatibility with major document classes, including \texttt{IEEEtran}, \texttt{acmart}, \texttt{iclr conference}, \texttt{article}, poster formats, and \texttt{beamer} presentations.

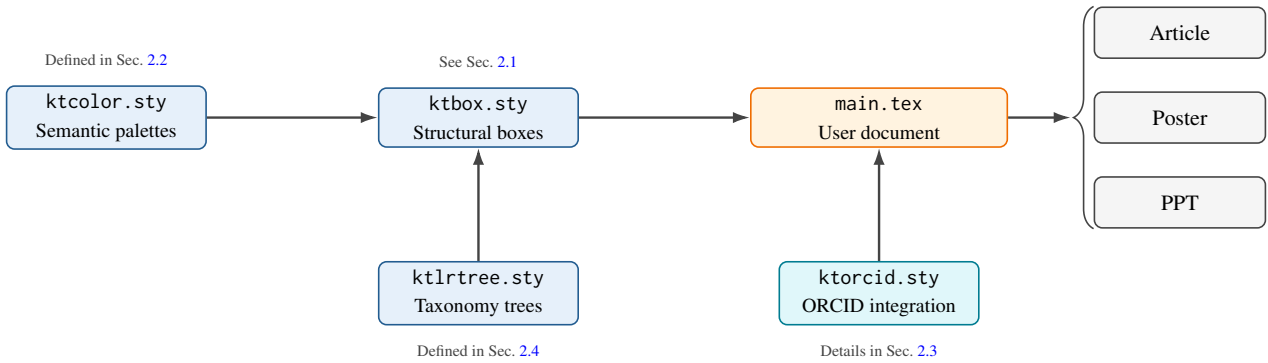
\begin{figure}[t]
  \centering
  \scalebox{0.75}{%
  \begin{tikzpicture}[>=latex, thick, font=\small]
    \usetikzlibrary{arrows.meta,decorations.pathreplacing}

    % Styles with ktcolor palette
    \tikzstyle{pkg}=[
      draw,
      rounded corners,
      minimum width=3.5cm,
      minimum height=1cm,
      align=center,
      fill=ktblue-bg,
      draw=ktblue-border
    ]
    \tikzstyle{main}=[
      draw,
      rounded corners,
      minimum width=4.5cm,
      minimum height=1cm,
      align=center,
      fill=ktorange-bg,
      draw=ktorange-border
    ]
    \tikzstyle{aux}=[
      draw,
      rounded corners,
      minimum width=3.5cm,
      minimum height=1cm,
      align=center,
      fill=ktcyan-bg,
      draw=ktcyan-border
    ]
    \tikzstyle{outnode}=[
      draw,
      rounded corners,
      minimum width=3cm,
      minimum height=0.9cm,
      align=center,
      fill=ktgray-bg,
      draw=ktgray-border
    ]
    \tikzstyle{arrow}=[-{Latex[length=3mm,width=2mm]}, very thick, draw=ktgray-border]

    % Nodes
    \node[pkg] (ktcolor) {\texttt{ktcolor.sty}\\{\footnotesize Semantic palettes}};
    \node[pkg, right=3cm of ktcolor] (ktbox) {\texttt{ktbox.sty}\\{\footnotesize Structural boxes}};
    \node[main, right=3cm of ktbox] (maintex) {\texttt{main.tex}\\{\footnotesize User document}};
    \node[aux, below=2cm of maintex] (ktorcid) {\texttt{ktorcid.sty}\\{\footnotesize ORCID integration}};
    \node[pkg, below=2cm of ktbox] (ktlrtree) {\texttt{ktlrtree.sty}\\{\footnotesize Taxonomy trees}};

    % Outputs
    \node[outnode, right=1.5cm of maintex, yshift=1.5cm] (article) {Article};
    \node[outnode, right=1.5cm of maintex] (poster) {Poster};
    \node[outnode, right=1.5cm of maintex, yshift=-1.5cm] (ppt) {PPT};

    % Arrows
    \draw[arrow] (ktcolor) -- (ktbox);
    \draw[arrow] (ktbox) -- (maintex);
    \draw[arrow] (ktorcid.north) -- (maintex.south);
    \draw[arrow] (maintex.east) -- ++(1.1,0);
    \draw[arrow] (ktlrtree.north) -- (ktbox.south);

    % Curly brace grouping outputs
    \draw[decorate,decoration={brace,amplitude=10pt},thick,draw=ktgray-border]
      (ppt.south west) -- (article.north west);

    % Annotations
    \node[align=left, text=ktgray-border, font=\scriptsize, above=0.2cm of ktcolor] {Defined in Sec.~\ref{sec:semantic-color-design}};
    \node[align=left, text=ktgray-border, font=\scriptsize, above=0.2cm of ktbox] {See Sec.~\ref{sec:structural-components}};
    \node[align=left, text=ktgray-border, font=\scriptsize, below=0.2cm of ktorcid] {Details in Sec.~\ref{sec:extendability-and-modularity}};
    \node[align=left, text=ktgray-border, font=\scriptsize, below=0.2cm of ktlrtree] {Defined in Sec.~\ref{sec:taxonomy-tree}};
  \end{tikzpicture}}
  \caption[ktbox framework workflow]{Dependency workflow of the \texttt{ktbox} framework, annotated with semantic color design, structural environments, taxonomy trees, and ORCID integration. Outputs (Article, Poster, PPT) are grouped under a unified brace.}
\end{figure}

\subsection{Structural Components}
\label{sec:structural-components}
  The structural design of the \texttt{ktbox} framework builds directly on the flexibility of the \texttt{tcolorbox} package. Rather than introducing new syntax or documentation overhead, the framework reuses the established key–value interface of \texttt{tcolorbox}, making the environments intuitive for LaTeX users who are already familiar with its ecosystem. This approach allows authors to immediately apply theming and layout keys without learning a separate layer of abstractions.

  Three core environments are defined: \texttt{ktbox}, \texttt{ktboxnumbered}, and \texttt{ktboxwide}. The standard \texttt{ktbox} provides a titled box with optional theming, suitable for concise highlights. The \texttt{ktboxnumbered} environment introduces automatic numbering for key takeaways, incrementing counters consistently across the document. This mechanism ensures structured presentation of recurring insights, while preserving compatibility with manual titles. In practice, it behaves as a hybrid between a highlight box and a titled note, aligning technical summaries with pedagogical clarity. Detailed usage examples are included in Appendix~\ref{sec:structural-numbered}.

  The \texttt{ktboxwide} variant extends across the page width, removing the bubble title and adopting a flat top style. This design is particularly suited for manuscripts where visual balance requires a minimal header or where wide content such as tables, enumerations, or code listings need to be integrated without misalignment. Since it inherits \texttt{tcolorbox}’s breakable behavior, page breaks are handled gracefully, while spacing adjustments can be achieved using \verb|\vspace| commands if additional separation from surrounding text is needed.

  A notable strength of the framework is its support for multi-column layouts. Both the standard and wide environments can be embedded in \texttt{minipage} structures, enabling two- and three-column arrangements for comparative displays. This feature allows authors to present metrics, figures, or parallel arguments with clarity, and is illustrated in Appendix~\ref{sec:structural-multicol}. The boxes also support nested environments, permitting the inclusion of tables, itemized lists, or codeblocks within a styled container.

  By extending \texttt{tcolorbox} natively rather than redefining its mechanics, the structural layer of the framework achieves a balance between flexibility and consistency. Authors can seamlessly combine highlight boxes, numbered insights, and wide-format containers, ensuring uniform design across diverse publication contexts without sacrificing the expressive range of LaTeX.

\subsection{Semantic Color Design}
\label{sec:semantic-color-design}
  The \texttt{ktcolor} package is designed around semantic mapping of colors rather than ad-hoc stylistic choices. Each palette defines roles such as \texttt{-bg} for background, \texttt{-title} for headings, \texttt{-border} for framing, and \texttt{-titlebox} for secondary contrast. This abstraction ensures that highlight environments maintain a coherent visual identity even when themes are changed or adapted for different publication contexts.

  \textbf{Choice of color codes.} Modern hex and RGB values are employed to guarantee consistency across digital and print outputs. Lighter palettes are based on high-contrast, pastel tones that maximize legibility on white paper or PDF backgrounds, while darker palettes use muted, desaturated hues to reduce glare during presentations. Colors are selected to balance saturation and luminance: bright shades are softened into pastels for readability (e.g., \#FFEBEE for \texttt{ktred-bg}), and darker backgrounds are complemented with lighter text tones for clarity (e.g., \#1F2A36 for \texttt{ktblue-bg-dark}).

  \textbf{Dual light and dark themes.} The framework explicitly supports both light and dark modes. Light themes are suited for printed manuscripts and PDF reading, where white backgrounds dominate. Dark themes address on-screen viewing in slides or posters, offering reduced brightness and visual comfort. The semantic mapping guarantees that a highlight box tagged as “key insight” retains its design role whether presented in light or dark mode.

  \textbf{Design aesthetics.} The visual philosophy emphasizes minimalism and consistency. Rounded corners, subtle shading, and neutral gray anchors are used to enhance readability without introducing clutter. Borders employ slightly darker shades of the primary hue, producing contrast while preserving harmony. This structure enables institutions, journals, or research groups to easily adopt or extend the palette to establish distinctive visual branding.

  By combining semantic abstraction, carefully curated color codes, dual-mode themeing, and minimalist aesthetics, \texttt{ktcolor} elevates color from a decorative choice into a structured component of scientific communication. This approach ensures that the \texttt{ktbox} framework remains accessible, visually clear, and reproducible across articles, posters, and presentations.

\subsection{Extendability and Modularity}
\label{sec:extendability-and-modularity}
  A defining property of the \texttt{ktbox} framework is its modular and scalable design, achieved through a strict separation of concerns. The color definitions in \texttt{ktcolor.sty} are fully independent of the structural components in \texttt{ktbox.sty}, which allows authors to extend or replace either layer without affecting the other. New color themes can be introduced seamlessly, while the \texttt{ktcolor} package may also be used in isolation when highlight boxes are not required, since it is a self-contained package depending only on \texttt{xcolor} (\verb|\RequirePackage[table,dvipsnames]{xcolor}|). At the same time, careful attention is paid to dependency management: although \texttt{tcolorbox} is not loaded by many standard \texttt{.cls} files, custom classes may introduce conflicting settings. The framework mitigates these risks by keeping theme definitions encapsulated in \texttt{ktcolor.sty} and invoking \texttt{tcolorbox} only where necessary. This separation of color palettes from structural logic ensures that the framework remains robust, portable, and straightforward to extend across articles, posters, and presentation contexts.

  \textbf{Independent utilities.} In addition to the structural and color components, the framework includes lightweight standalone modules such as \texttt{ktorcid}. This helper provides author metadata integration through ORCID identifiers, introducing two commands: \verb|\orcid| for rendering a full linked identifier and \verb|\orcidicon| for inserting a compact superscript icon. Since it depends only on \texttt{hyperref} and \texttt{orcidlink}, the module can be adopted independently of the rest of the framework, reinforcing the philosophy of modularity and ease of integration and is illustrated in Appendix~\ref{sec:orcid}.

\subsection{Taxonomy Tree (TT)}
\label{sec:taxonomy-tree}
  The motivation for introducing taxonomy trees in the ktbox framework emerges from a simple yet persistent challenge in academic writing where authors often need to communicate the breadth and evolution of methods through hierarchies but find themselves constrained by rigid or improvised formatting. In areas such as vDDD, where the literature spans early CNN architectures, temporal RNN and LSTM approaches, spatio temporal 3D CNNs, ensemble and attention based models, Vision Transformers, and most recently visual large language models, the complexity of representation demands a visual grammar that is both precise and aesthetically clear as illustrated in Fig.~\ref{fig:taxonomy-architectures}. Existing implementations with the forest package provide a foundation but the default top to bottom orientation, uniform node sizing, and limited styling make it difficult to achieve the intended clarity when constructing left to right hierarchies enriched with semantic cues and bibliographic references. The problem is compounded when deeper levels must be visually subordinated without breaking alignment, or when different semantic roles such as key takeaways and supporting variants must be differentiated consistently. The ktlrtree package addresses this gap by extending forest into a reusable component of the ktbox ecosystem, introducing dedicated styles for arrow based and plain linkages, seamless integration with the ktcolor semantic palettes, and helper macros for adjustable wrap boxes, curly brace groupings, and side by side fusion structures. These features allow complex taxonomies to be built incrementally with minimal effort while retaining a coherent visual identity across manuscripts, posters, and presentations. In practice this means that authors can present hierarchies with straight arrow linkages, distinct font sizes for different levels, and background colors that echo the semantic themes already present in highlight boxes, all while embedding citations directly inside nodes without disrupting layout. The result is a technically simple yet conceptually powerful tool that transforms taxonomic diagrams from ad hoc illustrations into reproducible scholarly assets aligned with the ktbox philosophy of semantic clarity and modular design, making complex structures approachable, human readable, and portable across diverse scholarly formats.

    %% Computer Vision Architectures - Generic Taxonomy Tree with Citations
    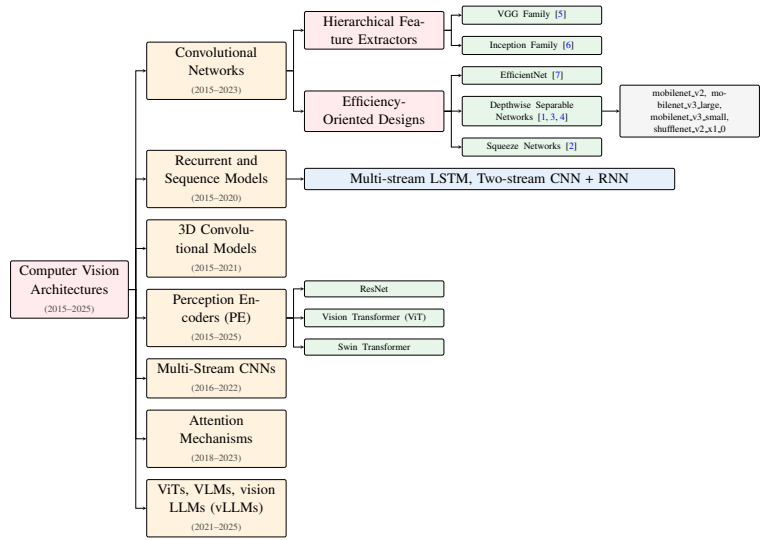
\begin{figure*}[h]
      \centering
      \scalebox{0.45}{
        \begin{forest} ktlrtree-arrow-unified
          [\ktwrapboxs{Computer Vision Architectures\\{\color{ktorange-bg-dark}\scriptsize(2015--2025)}}, fill=ktred-bg
            [\ktwrapboxm{Convolutional Networks\\{\color{ktorange-bg-dark}\scriptsize(2015--2023)}}, fill=ktorange-bg
              [\ktwrapboxm{Hierarchical Feature Extractors}, fill=ktred-bg
                [\ktwrapboxm{\scriptsize VGG Family~\cite{2015-simonyank-verydeepconvolutional}}, fill=ktgreen-bg]
                [\ktwrapboxm{\scriptsize Inception Family~\cite{2016-szegedyc-rethinkinginceptionarchitecture}}, fill=ktgreen-bg]
              ]
              [\ktwrapboxm{Efficiency-Oriented Designs}, fill=ktred-bg
                 [\ktwrapboxm{\scriptsize EfficientNet~\cite{2020-tanm-efficientnetrethinkingmodel}}, fill=ktgreen-bg]
                 [\ktwrapboxm{\scriptsize Depthwise Separable Networks~\cite{2019-sandlerm-mobilenetv2invertedresiduals,2019-howarda-searchingmobilenetv3,2018-man-shufflenetv2practical}}, fill=ktgreen-bg
                   [\ktwrapboxm{\scriptsize mobilenet\_v2, mobilenet\_v3\_large, mobilenet\_v3\_small, shufflenet\_v2\_x1\_0}, fill=ktgray-bg]
                 ]
                 [\ktwrapboxm{\scriptsize Squeeze Networks~\cite{2016-iandolafn-squeezenetalexnetlevelaccuracy}}, fill=ktgreen-bg]
               ]
            ]
            [\ktwrapboxm{Recurrent and Sequence Models\\{\color{ktorange-bg-dark}\scriptsize(2015--2020)}}, fill=ktorange-bg
              [\ktwrapboxxxxl{Multi-stream LSTM, Two-stream CNN + RNN}, fill=ktblue-bg]
            ]
            [\ktwrapboxm{3D Convolutional Models\\{\color{ktorange-bg-dark}\scriptsize(2015--2021)}}, fill=ktorange-bg]
            [\ktwrapboxm{Perception Encoders (PE)\\{\color{ktorange-bg-dark}\scriptsize(2015--2025)}}, fill=ktorange-bg
              [\ktwrapboxm{\scriptsize ResNet}, fill=ktgreen-bg]
              [\ktwrapboxm{\scriptsize Vision Transformer (ViT)}, fill=ktgreen-bg]
              [\ktwrapboxm{\scriptsize Swin Transformer}, fill=ktgreen-bg]
            ]
            [\ktwrapboxm{Multi-Stream CNNs\\{\color{ktorange-bg-dark}\scriptsize(2016--2022)}}, fill=ktorange-bg]
            [\ktwrapboxm{Attention Mechanisms\\{\color{ktorange-bg-dark}\scriptsize(2018--2023)}}, fill=ktorange-bg]
            [\ktwrapboxm{ViTs, VLMs, vision LLMs (vLLMs)\\{\color{ktorange-bg-dark}\scriptsize(2021--2025)}}, fill=ktorange-bg]
          ]
        \end{forest}
      }
      \caption[Computer Vision architectures]{Illustrative taxonomy tree (TT) for computer vision architectures. Each node highlights representative architectures with and without citations mix to show flexibility of TT.}
      \label{fig:taxonomy-architectures}
    \end{figure*}

  \noindent\textbf{Node sizing and wrapbox abstraction.} In ktlrtree, node widths are controlled through a graded family of wrap box commands that encapsulate \texttt{tcolorbox} into predefined sizes. Each macro corresponds to an explicit width: \verb|\ktwrapboxxs| (6em), \verb|\ktwrapboxs| (7.5em), \verb|\ktwrapboxm| (9em, default), \verb|\ktwrapboxl| (11em), \verb|\ktwrapboxxl| (13em), \verb|\ktwrapboxxxl| (15em), and \verb|\ktwrapboxxxxl| (25em). This scaling system is inspired by responsive design patterns where incremental labels (xs, s, m, xl) represent proportional steps rather than fixed semantics. Smaller variants are suited for concise labels or abbreviations, while larger ones accommodate extended names, multiple citations, or descriptive annotations. By abstracting widths into semantic commands, the framework ensures alignment and spacing remain consistent across levels of the taxonomy, freeing the author from low-level box configuration while preserving reproducibility.

  \noindent\textbf{Semantic alignment with themes.} The wrap box sizing operates in tandem with the ktcolor palettes, which define background, border, and title tones for different roles in the hierarchy. Together, they enable diagrams to highlight contrast between parent categories, intermediate families, and leaf-level instances without the clutter of manual color or font tuning. In practice, a taxonomy tree rendered with these elements reads like a structured visual narrative where scale, color, and spacing together convey both hierarchy and emphasis.

  \noindent\textbf{Conceptual overview.} The KTBox framework can itself be represented through a taxonomy, as illustrated in Fig.~\ref{fig:taxonomy-ktbox-framework}. The root node corresponds to the framework, branching into its principal modules: semantic color design (Sec.~\ref{sec:semantic-color-design}), structural boxes (Sec.~\ref{sec:structural-components}), taxonomy trees (Sec.~\ref{sec:taxonomy-tree}), and ORCID integration (App.~\ref{sec:orcid}). Each branch conveys the logical separation of concerns that underpins the framework design, while cross-references to sections reinforce reproducibility and traceability.  

%% Conceptual Representation of KTBox Framework
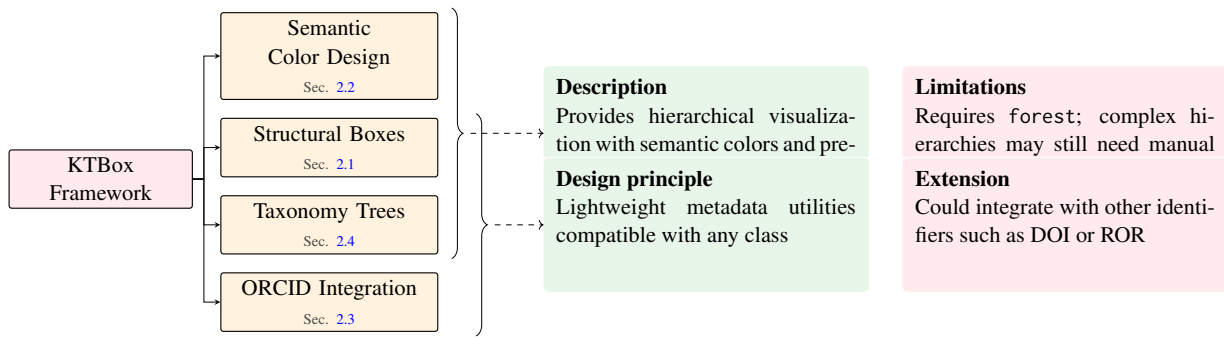
\begin{figure*}
  \centering
  \scalebox{0.7}{
    \begin{forest} ktlrtree-arrow-unified
      [\ktwrapboxs{KTBox Framework}, fill=ktred-bg
        [\ktwrapboxm{Semantic Color Design\\{\color{ktorange-bg-dark}\scriptsize Sec.~\ref{sec:semantic-color-design}}}, fill=ktorange-bg, name=a1]
        [\ktwrapboxm{Structural Boxes\\{\color{ktorange-bg-dark}\scriptsize Sec.~\ref{sec:structural-components}}}, fill=ktorange-bg, name=a2]
        [\ktwrapboxm{Taxonomy Trees\\{\color{ktorange-bg-dark}\scriptsize Sec.~\ref{sec:taxonomy-tree}}}, fill=ktorange-bg, name=a3,
          tikz+={
            \ktcurl{a123}{a1}{a3}{
              \ktfusionboxsplit
                {\textbf{Description}\\Provides hierarchical visualization with semantic colors and pre-sized nodes}
                {\textbf{Limitations}\\Requires \texttt{forest}; complex hierarchies may still need manual tuning}
            }
          }
        ]
        [\ktwrapboxm{ORCID Integration\\{\color{ktorange-bg-dark}\scriptsize Sec.~\ref{sec:extendability-and-modularity}}}, fill=ktorange-bg, name=a4,
          tikz+={
            \ktcurl[18pt]{a24}{a2}{a4}{
              \ktfusionboxsplit
                {\textbf{Design principle}\\Lightweight metadata utilities compatible with any class}
                {\textbf{Extension}\\Could integrate with other identifiers such as DOI or ROR}
            }
          }
        ]
      ]
    \end{forest}
  }
  \caption[Conceptual taxonomy of KTBox framework]{Conceptual taxonomy of the KTBox framework. Each module of the system is represented as a branch, with cross-references to the respective sections in the paper. Curly braces group related components, and the split boxes summarize their description and limitations or design principles and extensions. This representation mirrors the style of learning technique taxonomies but focuses on the internal architecture of the framework.}
  \label{fig:taxonomy-ktbox-framework}
\end{figure*}

  \noindent\textbf{Fusion helpers.} The diagram also demonstrates the use of fusion helpers provided in \texttt{ktlrtree}. The \verb|\ktcurl| macro draws a curly brace across a group of nodes and anchors an annotation box to the right, visually highlighting their collective role. Within this annotation, the \verb|\ktfusionboxsplit| helper creates a side-by-side comparison box. In Fig.~\ref{fig:taxonomy-ktbox-framework}, this mechanism is used to present \emph{description versus limitation} for taxonomy trees, and \emph{design principle versus extension} for ORCID integration. This structured annotation allows authors to extend hierarchical trees beyond static lists, embedding contextual commentary that maintains alignment with the visual grammar of the diagram.  

  \noindent\textbf{Integration within ktbox.} By combining hierarchical representation with semantic coloring and structured annotation, the conceptual taxonomy of KTBox illustrates how the same toolkit that styles research outputs can also be applied reflexively to describe its own architecture. The inclusion of braces and split boxes makes the diagram not merely a map of components but a compact narrative of their purpose, boundaries, and potential growth.

  %% Discussion and Future Work
  \section{Discussion and Future Work}
  \label{sec:discussion-and-future-work}
The modular philosophy of the \texttt{ktbox} framework opens up natural opportunities for extension beyond the current highlight boxes and palettes. A direct outcome of this principle has been the development of \texttt{ktlrtree}, a lightweight layer on top of the \texttt{forest} package that enables taxonomies to be rendered in a left-to-right layout with semantic colors, predefined node sizing, and fusion text boxes for annotated groupings. By abstracting away the low-level configuration of \texttt{forest}, \texttt{ktlrtree} allows hierarchical structures such as dataset families, model categories, or conceptual frameworks to be produced in a manner that is both aesthetically consistent and thematically aligned with the rest of the framework. In parallel, the idea of \texttt{kttables} emerges as a convenience wrapper for tabular design, allowing border, fill, and text colors to be applied thematically without requiring authors to manually embed color macros for each cell or rule. These extensions reinforce the central principle of abstraction: separating stylistic intent from implementation detail so that the same semantic layer that governs highlight boxes can seamlessly inform taxonomies and tables, thereby broadening the framework into a coherent ecosystem for structured, theme-aware scientific authoring.

  %% Conclusion
  \section{Conclusion}
  \label{sec:conclusion}
The \texttt{ktbox} framework establishes a principled approach to styling in scientific authoring by isolating color semantics, structural box logic, and auxiliary utilities into coherent modules that remain portable across diverse document classes. Its design ensures that highlight environments can be flexibly themed, scaled into numbered or wide variants, and integrated into multi-column layouts while maintaining consistency of appearance and behavior. With the addition of \texttt{ktlrtree}, hierarchical taxonomies can now be expressed in a left-to-right layout enriched with semantic colors, predefined node sizing, and fusion text boxes, further extending the framework beyond highlight boxes into diagrammatic representations. By keeping dependencies minimal and responsibilities clearly separated, the framework not only avoids conflicts with existing class files but also positions itself as an extensible foundation that can absorb future additions such as thematic tables. In doing so, it shifts color, structure, and hierarchy from ad-hoc embellishments into deliberate, reusable components of scholarly communication.

  %% Acknowledgment
  \section*{Acknowledgment}
This work was supported by C.E. Info Systems Ltd.\ (MapmyIndia) under the Industry Mentored Ph.D.\ in Advanced and Cutting-edge Technologies (PhD IMPACT) programme. The present contribution does not make use of proprietary code, data, or content from MapmyIndia. Instead, it extends auxiliary outcomes of the primary research activities, with the developed artifacts being released to the scientific community for broader use in scholarly communication.

  %% declaration
\section*{Large Language Models Usage}
This work used large language models (LLMs), including ChatGPT (OpenAI), for writing polish, coding assistance (formatting and reviews), and \LaTeX{} editing. All technical contributions and analyses were performed by the authors.

  %% References
  \bibliographystyle{plain}   % or abbrv, unsrt, alpha
  \bibliography{fig.taxonomy-cv-architectures}

  \clearpage
  \appendix
\section{User Guide and Illustrations}
\label{sec:user-guide}
  This appendix consolidates practical examples of the \texttt{ktbox} framework. It includes environment definitions, code snippets, and theme palettes. Each subsection corresponds to specific design and technical aspects described in Section~\ref{sec:framework}, allowing readers to directly connect conceptual discussions with concrete implementations. This document introduces custom key takeaway boxes. Each box supports optional `title' and `theme' parameters. Dark themes use the `-dark' suffix (e.g., `red-dark').

    \begin{itemize}
      \item \textbf{\texttt{ktbox}}: Standard titled box with theming.
      \item \textbf{\texttt{ktboxnumbered}}: Auto-numbered version for structured insights.
      \item \textbf{\texttt{ktboxwide}}: Full-width layout for high-emphasis content.
    \end{itemize}

  \subsection{Basic Highlight Box}
  \label{sec:basic-box}
    The fundamental environment for emphasis is the \texttt{ktbox}, which supports optional titles and theming.
    \vspace*{0.75em}

    \begin{ktbox}[title={Key Insight}]
      This is an example of a highlighted message.
    \end{ktbox}

    \begin{ktbox}[theme=gray]
      \begin{codeblock}
\vspace*{0.75em}
\begin{ktbox}[title={Key Insight}]
  This is an example of a highlighted message.
\end{ktbox}
      \end{codeblock}
    \end{ktbox}

  \subsection{Response Box Variation}
  \label{sec:response-box}
    The \texttt{ktbox} environment enables structured exchanges, such as author–reviewer dialogue.
    \vspace*{0.75em}
    \begin{ktbox}[title={Reviewer Comment}]
      This variation is designed for structured dialogue between authors and reviewers.
    \end{ktbox}

    \begin{ktbox}[theme=gray]
      \begin{codeblock}
\begin{ktbox}[title={Reviewer Comment}]
  This variation is designed for structured dialogue between authors and reviewers.
\end{ktbox}
      \end{codeblock}
    \end{ktbox}

  \subsection{Auto-Numbered Boxes}
  \label{sec:structural-numbered}
    The \texttt{ktboxnumbered} environment introduces built-in counters via \verb|\thetcbcounter|. This allows authors to present structured takeaways that are automatically numbered, ensuring consistency across sections. It behaves similarly to \texttt{ktbox}, except that the title argument is mandatory.

    \begin{ktboxnumbered}{Summary}
      Key observations are auto-numbered to improve traceability.
    \end{ktboxnumbered}

    \begin{ktbox}[theme=gray]
      \begin{codeblock}
\begin{ktboxnumbered}{Summary}
  Key observations are auto-numbered to improve traceability.
\end{ktboxnumbered}
      \end{codeblock}
    \end{ktbox}

  \subsection{Wide Layout Boxes}
  \label{sec:structural-wide}
    The \texttt{ktboxwide} environment omits the bubble title bar, making it ideal for manuscripts where seamless integration with body text is preferred. It expands to the full column width and avoids visual clutter while still retaining thematic consistency.

\begin{ktboxwide}[theme=orange]
  This wide box is best suited for papers without bubble titles.
\end{ktboxwide}

    \begin{ktbox}[theme=gray]
      \begin{codeblock}
\begin{ktboxwide}[theme=orange]
  This wide box is best suited for papers without bubble titles.
\end{ktboxwide}
      \end{codeblock}
    \end{ktbox}

  \subsection{Multi-Column Layouts}
  \label{sec:structural-multicol}
    By nesting boxes inside \texttt{minipage} environments wrapped in a \texttt{tcolorbox} container, the framework supports two- and three-column compositions. This is particularly useful for comparative metrics, side-by-side insights, or poster layouts.

    \textbf{Multi-Column Usage: Two Column Layout}
    \begin{tcolorbox}[enhanced, sharp corners=south, colframe=white, colback=white, boxrule=0pt, top=0pt, bottom=0pt, left=0pt, right=0pt]
      \begin{minipage}[t]{0.48\textwidth}
        \begin{ktbox}[title=Model Insight]
          MobileNetV3 offers a balanced trade-off for edge inference.
        \end{ktbox}
      \end{minipage}\hfill
      \begin{minipage}[t]{0.48\textwidth}
        \begin{ktbox}[theme=green, title=Camera Insight]
          Cam-1 and Cam-4 provide diverse modality views.
        \end{ktbox}
      \end{minipage}
    \end{tcolorbox}

    \begin{ktbox}[theme=gray]
      \begin{codeblock}
\begin{tcolorbox}[enhanced, sharp corners=south, colframe=white, colback=white, boxrule=0pt, top=0pt, bottom=0pt, left=0pt, right=0pt]
  \begin{minipage}[t]{0.48\textwidth}
    \begin{ktbox}[title=Model Insight]
      MobileNetV3 offers a balanced trade-off for edge inference.
    \end{ktbox}
  \end{minipage}\hfill
  \begin{minipage}[t]{0.48\textwidth}
    \begin{ktbox}[theme=green, title=Camera Insight]
      Cam-1 and Cam-4 provide diverse modality views.
    \end{ktbox}
  \end{minipage}
\end{tcolorbox}
      \end{codeblock}
    \end{ktbox}

    \textbf{Multi-Column Usage: Three Column Layout}
      \begin{tcolorbox}[enhanced, sharp corners=south, colframe=white, colback=white, boxrule=0pt, top=0pt, bottom=0pt, left=0pt, right=0pt]
        \begin{minipage}[t]{0.32\textwidth}
          \begin{ktbox}[theme=blue, title=Speed]
            MobileNetV2: \SI{165}{fps}
          \end{ktbox}
        \end{minipage}\hfill
        \begin{minipage}[t]{0.32\textwidth}
          \begin{ktbox}[theme=orange, title=Accuracy]
            ResNet50: \SI{91.3}{\percent}
          \end{ktbox}
        \end{minipage}\hfill
        \begin{minipage}[t]{0.32\textwidth}
          \begin{ktbox}[theme=red, title=Latency]
            Inference latency: \SI{13}{ms}
          \end{ktbox}
        \end{minipage}
      \end{tcolorbox}

      \begin{ktbox}[theme=gray]
        \begin{codeblock}
\begin{tcolorbox}[enhanced, sharp corners=south, colframe=white, colback=white, boxrule=0pt, top=0pt, bottom=0pt, left=0pt, right=0pt]
  \begin{minipage}[t]{0.32\textwidth}
    \begin{ktbox}[theme=blue, title=Speed]
      MobileNetV2: \SI{165}{fps}
    \end{ktbox}
  \end{minipage}\hfill
  \begin{minipage}[t]{0.32\textwidth}
    \begin{ktbox}[theme=orange, title=Accuracy]
      ResNet50: \SI{91.3}{\percent}
    \end{ktbox}
  \end{minipage}\hfill
  \begin{minipage}[t]{0.32\textwidth}
    \begin{ktbox}[theme=red, title=Latency]
      Inference latency: \SI{13}{ms}
    \end{ktbox}
  \end{minipage}
\end{tcolorbox}
        \end{codeblock}
      \end{ktbox}

  \subsection{Semantic Palettes}
  \label{sec:semantic-palettes}
    Examples of semantic palettes, as discussed in Section~\ref{sec:semantic-color-design}, are shown here. The demonstration includes variations defined in \texttt{ktcolor.sty}, illustrating how background, title, border, and titlebox roles are mapped.

  \subsection{Light and Dark Themes}
  \label{sec:light-dark}
    The framework explicitly supports both light and dark themes, enabling consistent appearance across print and on-screen settings. Tables~\ref{tab:light-themes} and \ref{tab:dark-themes} summarize the available palettes.

    \begin{table}[ht]
      \centering
      \renewcommand{\arraystretch}{1.15}
      \arrayrulecolor{ktblue-border}

      \begin{subtable}[t]{0.48\textwidth}
        \centering
        \caption{\textbf{Light Themes (Default)}}
        \label{tab:light-themes}

        \scriptsize
        \begin{tabular}{@{}lllll@{}}
          \toprule
          \rowcolor{ktblue-titlebox}
          \textcolor{ktblue-title}{\textbf{Theme}} &
          \textcolor{ktblue-title}{\textbf{Title}} &
          \textcolor{ktblue-title}{\textbf{Border}} &
          \textcolor{ktblue-title}{\textbf{TitleBox}} &
          \textcolor{ktblue-title}{\textbf{BG}} \\
          \midrule
          gray   & \cellcolor{ktgray-title}   & \cellcolor{ktgray-border}   & \cellcolor{ktgray-titlebox}   & \cellcolor{ktgray-bg} \\
          blue   & \cellcolor{ktblue-title}   & \cellcolor{ktblue-border}   & \cellcolor{ktblue-titlebox}   & \cellcolor{ktblue-bg} \\
          green  & \cellcolor{ktgreen-title}  & \cellcolor{ktgreen-border}  & \cellcolor{ktgreen-titlebox}  & \cellcolor{ktgreen-bg} \\
          yellow & \cellcolor{ktyellow-title} & \cellcolor{ktyellow-border} & \cellcolor{ktyellow-titlebox} & \cellcolor{ktyellow-bg} \\
          orange & \cellcolor{ktorange-title} & \cellcolor{ktorange-border} & \cellcolor{ktorange-titlebox} & \cellcolor{ktorange-bg} \\
          red    & \cellcolor{ktred-title}    & \cellcolor{ktred-border}    & \cellcolor{ktred-titlebox}    & \cellcolor{ktred-bg} \\
          \bottomrule
        \end{tabular}
      \end{subtable}
      \hfill
      \begin{subtable}[t]{0.48\textwidth}
        \centering
        \caption{\textbf{Dark Mode Themes}}
        \label{tab:dark-themes}
        \scriptsize
        \begin{tabular}{@{}llllll@{}}
          \toprule
          \rowcolor{ktblue-titlebox}
          \textcolor{ktblue-title}{\textbf{Theme}} &
          \textcolor{ktblue-title}{\textbf{Title}} &
          \textcolor{ktblue-title}{\textbf{Border}} &
          \textcolor{ktblue-title}{\textbf{TitleBox}} &
          \textcolor{ktblue-title}{\textbf{Text}} &
          \textcolor{ktblue-title}{\textbf{BG}} \\
          \midrule
          gray   & \cellcolor{ktgray-title-dark}   & \cellcolor{ktgray-border-dark}   & \cellcolor{ktgray-titlebox-dark}   & \cellcolor{ktgray-text-dark}   & \cellcolor{ktgray-bg-dark} \\
          blue   & \cellcolor{ktblue-title-dark}   & \cellcolor{ktblue-border-dark}   & \cellcolor{ktblue-titlebox-dark}   & \cellcolor{ktblue-text-dark}   & \cellcolor{ktblue-bg-dark} \\
          green  & \cellcolor{ktgreen-title-dark}  & \cellcolor{ktgreen-border-dark}  & \cellcolor{ktgreen-titlebox-dark}  & \cellcolor{ktgreen-text-dark}  & \cellcolor{ktgreen-bg-dark} \\
          yellow & \cellcolor{ktyellow-title-dark} & \cellcolor{ktyellow-border-dark} & \cellcolor{ktyellow-titlebox-dark} & \cellcolor{ktyellow-text-dark} & \cellcolor{ktyellow-bg-dark} \\
          orange & \cellcolor{ktorange-title-dark} & \cellcolor{ktorange-border-dark} & \cellcolor{ktorange-titlebox-dark} & \cellcolor{ktorange-text-dark} & \cellcolor{ktorange-bg-dark} \\
          red    & \cellcolor{ktred-title-dark}    & \cellcolor{ktred-border-dark}    & \cellcolor{ktred-titlebox-dark}    & \cellcolor{ktred-text-dark}    & \cellcolor{ktred-bg-dark} \\
          \bottomrule
        \end{tabular}
      \end{subtable}

      \arrayrulecolor{black}
    \end{table}

  \subsection{Design Aesthetics and Layouts}
  \label{sec:design-aesthetics}
    Illustrations of minimalism in layout, use of rounded corners, and multi-column arrangements are provided here. These reinforce the design philosophy described in Section~\ref{sec:semantic-color-design}.

  \subsection{Highlight Boxes}
  \label{sec:structural-boxes}
    The core environments for emphasis are built on \texttt{tcolorbox}, extended with semantic theming. The standard \texttt{ktbox} supports both optional titles and themes, making it the most versatile choice for inline highlights.

    \begin{ktbox}[theme=gray]
\begin{codeblock}
  \begin{ktbox}[title={Model Insight}, theme=green]
  This box combines a title with semantic theming.
  \end{ktbox}
\end{codeblock}
    \end{ktbox}

  \subsection{With Theme}
  \label{with-theme}
    \vspace*{0.75em}

    \begin{ktboxnumbered}[theme=red]{Insight}
      MobileNetV3 achieves optimal real-time performance for embedded systems.
    \end{ktboxnumbered}

    \begin{ktboxnumbered}[theme=orange]{Insight}
      MobileNetV3 achieves optimal real-time performance for embedded systems.
    \end{ktboxnumbered}

    \begin{ktboxnumbered}[theme=green]{Insight}
      MobileNetV3 achieves optimal real-time performance for embedded systems.
    \end{ktboxnumbered}

    \begin{ktboxnumbered}[theme=yellow]{Insight}
      MobileNetV3 achieves optimal real-time performance for embedded systems.
    \end{ktboxnumbered}

    \begin{ktboxnumbered}[theme=blue]{Insight}
      MobileNetV3 achieves optimal real-time performance for embedded systems.
    \end{ktboxnumbered}

    \begin{ktboxnumbered}[theme=cyan]{Insight}
      MobileNetV3 achieves optimal real-time performance for embedded systems.
    \end{ktboxnumbered}

    \begin{ktboxnumbered}[theme=purple]{Insight}
      MobileNetV3 achieves optimal real-time performance for embedded systems.
    \end{ktboxnumbered}

    \begin{ktboxnumbered}[theme=magenta]{Insight}
      MobileNetV3 achieves optimal real-time performance for embedded systems.
    \end{ktboxnumbered}

    \begin{ktboxnumbered}[theme=gray]{Insight}
      MobileNetV3 achieves optimal real-time performance for embedded systems.
    \end{ktboxnumbered}

    \begin{ktboxnumbered}[theme=white]{Insight}
      MobileNetV3 achieves optimal real-time performance for embedded systems.
    \end{ktboxnumbered}

  \subsection{ORCID Integration}
  \label{sec:orcid}
    The \texttt{ktorcid} package extends author metadata by embedding ORCID identifiers with icons.

    \begin{ktbox}[theme=gray]
\begin{codeblock}
  \author{Bhaskar Mangal\orcidicon{0000-0002-8126-3528}}
\end{codeblock}
    \end{ktbox}

  \subsection{Taxonomy Tree Layouts}
  \label{sec:taxonomy-tree-layouts}
    The appendix subsection in Fig.~\ref{fig:taxonomy-architectures} provides a code example that demonstrates how the taxonomy tree layouts introduced in the main text can be directly implemented using the \texttt{ktlrtree} extension of the \texttt{forest} package. The snippet illustrates a left to right hierarchical organization of computer vision architectures, beginning with broad categories such as convolutional networks, recurrent and sequence models, three dimensional convolutions, perception encoders, and attention based designs, and then progressively refining these into well known model families including VGG~\cite{2015-simonyank-verydeepconvolutional}, Inception~\cite{2016-szegedyc-rethinkinginceptionarchitecture}, EfficientNet~\cite{2020-tanm-efficientnetrethinkingmodel}, MobileNet~\cite{2019-sandlerm-mobilenetv2invertedresiduals,2019-howarda-searchingmobilenetv3}, and SqueezeNet~\cite{2016-iandolafn-squeezenetalexnetlevelaccuracy}. The layout makes use of semantic coloring defined by the ktbox framework to visually distinguish parent and child nodes, while helper commands such as \verb|\ktwrapboxm| and \verb|\ktwrapboxxxxl| ensure that nodes are appropriately sized and aligned for readability. By embedding citation keys directly within nodes, the structure preserves scholarly attribution while maintaining an aesthetically coherent diagram. This example highlights how the declarative design of ktlrtree simplifies the otherwise complex task of aligning edges, adjusting node sizes, and balancing fonts in forest diagrams, and the code block shown in the appendix mirrors the rendered figure presented in the main content, offering readers a reproducible reference for creating their own taxonomy trees.

    \begin{ktbox}[theme=gray]
\begin{codeblock}
  %% Computer Vision Architectures - Generic Taxonomy Tree with Citations
  \begin{figure*}[h]
    \centering
    \scalebox{0.45}{
      \begin{forest} ktlrtree-arrow-unified
        [\ktwrapboxs{Computer Vision Architectures\\{\color{ktorange-bg-dark}\scriptsize(2015--2025)}}, fill=ktred-bg
          [\ktwrapboxm{Convolutional Networks\\{\color{ktorange-bg-dark}\scriptsize(2015--2023)}}, fill=ktorange-bg
            [\ktwrapboxm{Hierarchical Feature Extractors}, fill=ktred-bg
              [\ktwrapboxm{\scriptsize VGG Family~\cite{2015-simonyank-verydeepconvolutional}}, fill=ktgreen-bg]
              [\ktwrapboxm{\scriptsize Inception Family~\cite{2016-szegedyc-rethinkinginceptionarchitecture}}, fill=ktgreen-bg]
            ]
            [\ktwrapboxm{Efficiency-Oriented Designs}, fill=ktred-bg
               [\ktwrapboxm{\scriptsize EfficientNet~\cite{2020-tanm-efficientnetrethinkingmodel}}, fill=ktgreen-bg]
               [\ktwrapboxm{\scriptsize Depthwise Separable Networks~\cite{2019-sandlerm-mobilenetv2invertedresiduals,2019-howarda-searchingmobilenetv3,2018-man-shufflenetv2practical}}, fill=ktgreen-bg
                 [\ktwrapboxm{\scriptsize mobilenet\_v2, mobilenet\_v3\_large, mobilenet\_v3\_small, shufflenet\_v2\_x1\_0}, fill=ktgray-bg]
               ]
               [\ktwrapboxm{\scriptsize Squeeze Networks~\cite{2016-iandolafn-squeezenetalexnetlevelaccuracy}}, fill=ktgreen-bg]
             ]
          ]
          [\ktwrapboxm{Recurrent and Sequence Models\\{\color{ktorange-bg-dark}\scriptsize(2015--2020)}}, fill=ktorange-bg
            [\ktwrapboxxxxl{Multi-stream LSTM, Two-stream CNN + RNN}, fill=ktblue-bg]
          ]
          [\ktwrapboxm{3D Convolutional Models\\{\color{ktorange-bg-dark}\scriptsize(2015--2021)}}, fill=ktorange-bg]
          [\ktwrapboxm{Perception Encoders (PE)\\{\color{ktorange-bg-dark}\scriptsize(2015--2025)}}, fill=ktorange-bg
            [\ktwrapboxm{\scriptsize ResNet}, fill=ktgreen-bg]
            [\ktwrapboxm{\scriptsize Vision Transformer (ViT)}, fill=ktgreen-bg]
            [\ktwrapboxm{\scriptsize Swin Transformer}, fill=ktgreen-bg]
          ]
          [\ktwrapboxm{Multi-Stream CNNs\\{\color{ktorange-bg-dark}\scriptsize(2016--2022)}}, fill=ktorange-bg]
          [\ktwrapboxm{Attention Mechanisms\\{\color{ktorange-bg-dark}\scriptsize(2018--2023)}}, fill=ktorange-bg]
          [\ktwrapboxm{ViTs, VLMs, vision LLMs (vLLMs)\\{\color{ktorange-bg-dark}\scriptsize(2021--2025)}}, fill=ktorange-bg]
        ]
      \end{forest}
    }
    \caption[Computer Vision architectures]{Computer Vision architectures}
    \label{fig:taxonomy-architectures}
  \end{figure*}
\end{codeblock}
    \end{ktbox}

  \subsection{Taxonomy Tree Layouts with Fusion Text Box}
  \label{sec:taxonomy-tree-layouts-with-fusion-text-box}
    The listing in this subsection provides the complete \LaTeX\ code for the conceptual taxonomy of the KTBox framework illustrated earlier in Fig.~\ref{fig:taxonomy-ktbox-framework}. It reproduces the diagram by combining the predefined node sizing commands with the curly brace and fusion text box helpers, offering a reference implementation that readers can adapt directly in their own documents.

    \begin{ktbox}[theme=gray]
\begin{codeblock}
%% Conceptual Representation of KTBox Framework
\begin{figure*}
  \centering
  \scalebox{0.7}{
    \begin{forest} ktlrtree-arrow-unified
      [\ktwrapboxs{KTBox Framework}, fill=ktred-bg
        [\ktwrapboxm{Semantic Color Design\\{\color{ktorange-bg-dark}\scriptsize Sec.~\ref{sec:semantic-color-design}}}, fill=ktorange-bg, name=a1]
        [\ktwrapboxm{Structural Boxes\\{\color{ktorange-bg-dark}\scriptsize Sec.~\ref{sec:structural-components}}}, fill=ktorange-bg, name=a2]
        [\ktwrapboxm{Taxonomy Trees\\{\color{ktorange-bg-dark}\scriptsize Sec.~\ref{sec:taxonomy-tree}}}, fill=ktorange-bg, name=a3,
          tikz+={
            \ktcurl{a123}{a1}{a3}{
              \ktfusionboxsplit
                {\textbf{Description}\\Provides hierarchical visualization with semantic colors and pre-sized nodes}
                {\textbf{Limitations}\\Requires \texttt{forest}; complex hierarchies may still need manual tuning}
            }
          }
        ]
        [\ktwrapboxm{ORCID Integration\\{\color{ktorange-bg-dark}\scriptsize Sec.~\ref{sec:extendability-and-modularity}}}, fill=ktorange-bg, name=a4,
          tikz+={
            \ktcurl[18pt]{a24}{a2}{a4}{
              \ktfusionboxsplit
                {\textbf{Design principle}\\Lightweight metadata utilities compatible with any class}
                {\textbf{Extension}\\Could integrate with other identifiers such as DOI or ROR}
            }
          }
        ]
      ]
    \end{forest}
  }
  \caption[Conceptual taxonomy of KTBox framework]{Conceptual taxonomy of the KTBox framework. Each module of the system is represented as a branch, with cross-references to the respective sections in the paper. Curly braces group related components, and the split boxes summarize their description and limitations or design principles and extensions. This representation mirrors the style of learning technique taxonomies but focuses on the internal architecture of the framework.}
  \label{fig:taxonomy-ktbox-framework}
\end{figure*}
\end{codeblock}
    \end{ktbox}

\end{document}